\documentclass[12pt]{article}
\begin{document}
\begin{titlepage}
\begin{flushright}
hep-th/0511241\\
TIT/HEP-547\\
November, 2005\\
\end{flushright}
\vspace{0.5cm}
\begin{center}
{\Large \bf 
Central Charges in 
Non(anti)commutative ${\cal N}=2$ Supersymmetric $U(N)$ Gauge Theory
}
\lineskip .75em
\vskip2.5cm
{\large Katsushi Ito and Hiroaki Nakajima}
\vskip 2.5em
{\large\it Department of Physics\\
Tokyo Institute of Technology\\
Tokyo, 152-8551, Japan}  \vskip 4.5em
\end{center}
\begin{abstract}
We study the central charge of the  deformed ${\cal N}=(1,0)$ 
supersymmetry algebra in non(anti)commutative 
${\cal N}=2$ supersymmetric $U(N)$ gauge theory. 
In the cases of ${\cal N}=1/2$ superspace 
and ${\cal N}=2$ harmonic superspace with the singlet deformation, 
we find that the central charge is deformed by the non(anti)commutative 
parameters but depends on the electric and magnetic charges. 
For generic deformation of ${\cal N}=2$ harmonic superspace, 
we compute the $O(C)$ correction to the 
central charges in the case of $U(1)$ gauge group.

\end{abstract}
\end{titlepage}

\baselineskip=0.7cm
Supersymmetric field theories in non(anti)commutative superspace \cite{ScNi,Se}
arise as the low-energy effective field theories on 
 the D-branes in the graviphoton background \cite{OoVa,BeSe,DeGrNi} 
and have been extensively studied in the last few years.
These theories are defined in Euclidean superspace by using the 
$*$-product for the supercoordinates.
In particular supersymmetric field theories in  non(anti)commutative
${\cal N}=1$ superspace (${\cal N}=1/2$ superspace) 
have been investigated both perturbatively
and nonperturbatively \cite{ArItOh1,pert,inst,sigma}.

In the case of 
${\cal N}=2$ superspace, there exists a variety of deformations 
\cite{KlPeTa,FeSo,IvLeZu,ArItOh2}.
For generic $Q$-deformation, ${\cal N}=(1,1)$ supersymmetry is deformed 
to ${\cal N}=(1,0)$ \cite{IvLeZu,ArItOh2}. 
But for some particular deformation parameters, in
which only ${\cal N}=1$ subspace is deformed, it is shown that
the ${\cal N}=(1,0)$ supersymmetry enhances to  ${\cal N}=(1,1/2)$
\cite{IvLeZu}. 
The deformed ${\cal N}=(1,1/2)$ supersymmetry has been constructed 
explicitly in \cite{ArItOh5} for the $U(1)$ gauge theory.

For ${\cal N}=2$ supersymmetric $U(N)$ gauge theory in
non(anti)commutative ${\cal N}=1$ superspace, the deformed
${\cal N}=(1,1/2)$ supersymmetry is constructed in \cite{ItNa1}.
It is interesting to study the role of the
 deformed supersymmetry at the quantum level.
Since non(anti)commutative field theories do not have 
Poincar\'e invariance, the supersymmetry algebra could get nontrivial
corrections,  which avoid the Haag-{\L}opusza\'{n}ski-Sohnius
no-go theorem \cite{HaLoSo}. 
In ${\cal N}=1/2$ supersymmetric field theories, the central charge 
of the deformed supersymmetry algebra was studied in \cite{BPS,ChIn}.
In particular, for ${\cal N}=1/2$ Wess-Zumino (WZ) model, 
it was shown that the formula of the central charge 
associated with the domain wall is not deformed but the
non(anti)commutative effects enter through the deformed equations of
motion\cite{ChIn}.

In ${\cal N}=2$ supersymmetric gauge theory, 
Witten and Olive
\cite{WiOl} have shown that there exists a central extension in ${\cal N}=2$
supersymmetry algebra and its central charge is related to the electric
and magnetic charges of the monopoles or dyons of the theory
(See \cite{AlHa} for reviews).
Their BPS property is very important to study the strong coupling physics
of the theory \cite{SeWi}.

In this paper we will study the effects of non(anti)commutativity on the
deformed ${\cal N}=(1,0)$ supersymmetry algebra
of ${\cal N}=2$ supersymmetric $U(N)$ gauge theory  in 
${\cal N}=1/2$ superspace.
We find that new central charges appear in the algebra
but they depend on the electric
and magnetic charges and the vacuum expectation value of the Higgs fields.
We then extend this result to the non(anti)commutative ${\cal N}=2$ 
harmonic superspace.
Since the exact $U(N)$ action is only known for the singlet deformation
\cite{FeIvLeSoZu}, we calculate the ${\cal N}=(1,0)$ algebra based on this 
theory.
For generic deformation, the $O(C)$ $U(1)$ action has been obtained in
\cite{ArItOh2}. We will calculate the $O(C)$ 
correction to the central charge.

We begin with reviewing 
${\cal N}=2$ supersymmetric $U(N)$ gauge theory in
non(anti)com\-mut\-at\-ive ${\cal N}=1$ superspace.
Let $(x^{\mu},\theta^{\alpha},\bar{\theta}^{\dot{\alpha}})$ 
($\mu=0,\ldots,3$, $\alpha,\dot{\alpha}=1,2$) be 
supercoordinates of ${\cal N}=1$  Euclidean superspace 
and $\sigma^{\mu}_{\alpha\dot{\alpha}}$ and 
$\bar{\sigma}^{\mu\dot{\alpha}\alpha}$ Dirac matrices \cite{WeBa}.
We note that in Euclidean spacetime chiral and antichiral 
fermions transform independently under the Lorentz transformations.
$Q_\alpha={\partial\over \partial\theta^\alpha}
-i\sigma^{\mu}_{\alpha\dot{\alpha}}
\bar{\theta}^{\dot{\alpha}}\partial_{\mu}$ and
$\bar{Q}^{\dot{\alpha}}=
-{\partial\over \partial\bar{\theta}_{\dot{\alpha}}}+i
\theta_{\alpha}\bar{\sigma}^{\mu\dot{\alpha}\alpha}
\partial_{\mu}$ are supercharges.
$D_\alpha={\partial\over \partial\theta^\alpha}
+i\sigma^{\mu}_{\alpha\dot{\alpha}}
\bar{\theta}^{\dot{\alpha}}
\partial_{\mu}$
and
$\bar{D}_{\dot{\alpha}}=
-{\partial\over \partial\bar{\theta}_{\dot{\alpha}}}-i
\theta_{\alpha}\bar{\sigma}^{\mu\dot{\alpha}\alpha}
\partial_{\mu}$ are the supercovariant derivatives.
$\sigma^{\mu\nu}={1\over4}(\sigma^{\mu}\bar{\sigma}^{\nu}
-\sigma^{\nu}\bar{\sigma}^{\mu})$
and 
$\bar{\sigma}^{\mu\nu}={1\over4}(\bar{\sigma}^{\mu}\sigma^{\nu}
-\bar{\sigma}^{\nu}\sigma^{\mu})$
are the Lorentz generators.

The non(anti)commutativity in ${\cal N}=1$ superspace is introduced by 
the $*$-product:
\begin{equation}
 f*g(x,\theta,\bar{\theta})=f(x,\theta,\bar{\theta})
\exp\left(-{1\over2}\overleftarrow{Q}_\alpha C^{\alpha\beta}
\overrightarrow{Q}_\beta \right)g(x,\theta,\bar{\theta}).
\end{equation}
Using this $*$-product, the anticommutation relations for 
$\theta$ become
\begin{eqnarray}
 \left\{\theta^\alpha, \theta^\beta\right\}_{*}&=& C^{\alpha\beta}
\end{eqnarray}
while the chiral coordinates $y^{\mu}=x^{\mu}+i\theta\sigma^{\mu}\bar{\theta}$ 
and $\bar{\theta}$ are still commuting and anticommuting coordinates, 
respectively.

${\cal N}=2$ supersymmetric $U(N)$ gauge theory in this deformed
superspace was formulated in \cite{ArItOh1}.
It can be constructed by 
vector superfields $V$, chiral superfields $\Phi$ and an anti-chiral
superfields
$\bar{\Phi}$, where $\Phi$ and $\bar{\Phi}$ belong to the adjoint 
representation of $U(N)$.
We introduce the basis $t^a$ ($a=1,\cdots, N^2$) of the Lie algebra of $U(N)$,
normalized as ${\rm tr}(t^a t^b)=k\delta^{ab}$. 
The Lagrangian is 
\begin{eqnarray}
{\cal L}
={1\over k}
\int d^2\theta d^2\bar{\theta}\,
{\rm tr}(\bar{\Phi}*e^{V}*\Phi* e^{-V})
+{1\over 16kg^2}{\rm tr}\left(\int d^2\theta
 W^{\alpha}*W_{\alpha}
+\int d^2\bar{\theta}\bar{W}_{\dot{\alpha}}*\bar{W}^{\dot{\alpha}}
\right),
\label{eq:lag1}
\end{eqnarray}
where $g$ denotes the coupling constant.
$W_\alpha = -{1\over4}\bar{D}^2e^{-V}D_\alpha e^{V}$ 
and $\bar{W}_{\dot{\alpha}}={1\over 4}D^2
e^{-V}\bar{D}_{\dot{\alpha}}e^{V}$
are the chiral and antichiral field strengths.
Note that multiplication of superfields are defined by the $*$-product.
This Lagrangian is invariant under the gauge transformations
$\Phi\rightarrow e^{-i\Lambda}*\Phi* e^{i\Lambda}$, 
$\bar{\Phi}\rightarrow e^{-i\bar{\Lambda}}*\bar{\Phi}*e^{i\bar{\Lambda}}$
and $e^{V}\rightarrow e^{-i\bar{\Lambda}}*e^{V}*e^{i\Lambda}$.
To write down the Lagrangian in terms of component fields, 
it is convenient to take the WZ gauge as in the commutative case.
Since the $*$-product deforms the gauge transformation,
it is necessary to redefine the component fields such that 
these transform canonically under the gauge
transformation\cite{Se,ArItOh1}.
For ${\cal N}=2$ $U(N)$ theory, these superfields in the WZ gauge are 
\begin{eqnarray}
 \Phi(y,\theta)&=& A(y)+\sqrt{2}\theta\psi(y)+\theta\theta F(y),
\nonumber\\
\bar{\Phi}(\bar{y},\bar{\theta})&=&
\bar{A}(\bar{y})+\sqrt{2}\bar{\theta}\bar{\psi}(\bar{y})
+\bar{\theta}\bar{\theta}\left(
\bar{F}+i C^{\mu\nu}\partial_{\mu}\left\{ v_{\nu},\bar{A}\right\}
-{1\over4}C^{\mu\nu}\left[ v_{\mu},\left\{ v_{\nu},\bar{A}\right\}\right]
\right)(\bar{y}),
\nonumber\\
 V(y,\theta,\bar{\theta})&=&
-\theta \sigma^{\mu}\bar{\theta} v_{\mu}(y)
+i \theta\theta \bar{\theta}\bar{\lambda}(y)
-i \bar{\theta}\bar{\theta}\theta^{\alpha}
\left(\lambda_\alpha
+{1\over4}\varepsilon_{\alpha\beta}C^{\beta\gamma}
\left\{(\sigma^{\mu}\bar{\lambda})_{\gamma}, v_{\mu}\right\}
\right)(y)
\nonumber\\
&& +{1\over2} \theta\theta \bar{\theta}\bar{\theta}
(D-i\partial^{\mu}v_{\mu})(y).
\label{eq:wz1}
\end{eqnarray}
Here $\bar{y}^{\mu}=x^{\mu}-i\theta\sigma^{\mu}\bar{\theta}$ are the antichiral
coordinates
and
$C^{\mu\nu}=C^{\alpha\beta}\varepsilon_{\beta\gamma}(\sigma^{\mu\nu})_{\alpha}
{}^{\gamma}$.
Since $\sigma^{\mu\nu}$ is self-dual, $C^{\mu\nu}$ is also self-dual.
Substituting (\ref{eq:wz1}) into the Lagrangian (\ref{eq:lag1}), we
obtain
the deformed Lagrangian written in
terms of component fields.
In this expression, however, normalizations of 
two fermions $\psi$ and $\lambda$ are different.
In order to see symmetries between two fermions manifestly,
it is useful to rescale $V$ to $2g V$ and $C^{\alpha\beta}$ to ${1\over
2g}C^{\alpha\beta}$. 
Then the Lagrangian takes the form 
\begin{eqnarray}
 {\cal L}&=&
\frac{1}{k}{\rm tr}\Bigl(
-\frac{1}{4}F^{\mu\nu}F_{\mu\nu}-\frac{1}{4}F^{\mu\nu}\tilde{F}_{\mu\nu}
-i \bar{\lambda}\bar{\sigma}^{\mu}D_{\mu}\lambda
+\frac{1}{2}\tilde{D}^2\nonumber\\
&&
\!\!\!\!\!\!\!\!\!\!\!\!
-(D^{\mu}\bar{A})D_{\mu}A
-i \bar{\psi}\bar{\sigma}^{\mu}D_{\mu}\psi
+\bar{F} F
-i\sqrt{2} g[\bar{A},\psi]\lambda
-i\sqrt{2} g[A,\bar{\psi}]\bar{\lambda}
-{g^2\over2}[A,\bar{A}]^2
\Bigr)\nonumber\\
&&
+\frac{1}{k} {\rm tr} 
\left( - \frac{i}{2} C^{\mu\nu} F_{\mu\nu} \bar{\lambda} 
\bar{\lambda} 
+ \frac{1}{8} |C|^2 (\bar{\lambda} \bar{\lambda})^2 \right. 
\nonumber\\
& & \left. +\frac{i}{2} C^{\mu\nu} F_{\mu\nu} \{ \bar{A},F \} 
- \frac{\sqrt{2}}{2} C^{\alpha \beta} \{ D_{\mu} \bar{A} , 
(\sigma^{\mu} \bar{\lambda})_{\alpha} \} \psi_{\beta}  
- \frac{1}{16} |C|^2 [ \bar{A} , \bar{\lambda}] [\bar{\lambda} , F] 
\right), 
\label{eq:lag2a}
\end{eqnarray}
where $F_{\mu\nu}=\partial_{\mu} v_{\nu}-\partial_{\nu} v_{\mu}+ig [v_{\mu},v_{\nu}]$, 
$\tilde{F}_{\mu\nu}=\frac{1}{2}\epsilon_{\mu\nu\rho\sigma}F^{\rho\sigma}$, 
$|C|^2=C^{\mu\nu}C_{\mu\nu}$ and
$D_{\mu} \lambda=\partial_{\mu} \lambda+ig [v_{\mu}, \lambda]$ etc.
We have also introduced an auxiliary field $\tilde{D}$ defined by
$\tilde{D}=D+g [A,\bar{A}]$ in order to see undeformed ${\cal N}=2$
supersymmetry in a symmetric way.
It is shown in \cite{ArItOh1, ItNa1} 
that the action is invariant under the 
deformed ${\cal N}=(1,1/2)$ supersymmetry
\begin{eqnarray}
 \delta_{\xi}v_{\mu}&=& i \xi \sigma_{\mu}\bar{\lambda}, \nonumber\\
 \delta_{\xi}\lambda_\alpha&=& i\xi_\alpha \tilde{D}-ig\xi_\alpha [A,\bar{A}]
+(\sigma^{\mu\nu}\xi)_\alpha
\left( F_{\mu\nu}
+{i\over2}C_{\mu\nu}\bar{\lambda}\bar{\lambda}
\right),
\quad
 \delta_\xi\bar{\lambda}=0, \nonumber\\
 \delta_\xi \tilde{D}&=& -\xi \sigma^{\mu} D_{\mu}\bar{\lambda}
+\sqrt{2}g[\xi\psi,\bar{A}],
\nonumber\\
\delta_\xi A&=& \sqrt{2}\xi\psi,
\quad 
\delta_\xi \psi= \sqrt{2}\xi F, 
\quad
\delta_\xi F=0, 
\nonumber\\
\delta_\xi \bar{A}&=&0,
\nonumber\\
\delta_\xi\bar{\psi}&=& \sqrt{2}i \bar{\sigma}^{\mu}\xi D_{\mu}\bar{A},
\nonumber\\
\delta_\xi \bar{F}&=& i\sqrt{2}\xi\sigma^{\mu} D_{\mu}\bar{\psi}
-2gi [\bar{A},\xi\lambda]
+C^{\mu\nu}D_{\mu}\left\{ \bar{A},\xi\sigma_{\nu}\bar{\lambda}\right\},
\label{eq:tra1}
\end{eqnarray}
\begin{eqnarray}
 \delta_\eta v_{\mu}&=& -i \eta \sigma_{\mu}\bar{\psi}
-{\sqrt{2}\over2}C^{\alpha\beta}\eta_\alpha\left\{ \bar{A}, 
(\sigma_{\mu}\bar{\lambda})_{\beta}\right\},
\nonumber\\
 \delta_\eta \lambda^\alpha&=& \sqrt{2}\eta^\alpha \bar{F}
\nonumber\\
&&
-{\sqrt{2}\over2}C^{\alpha\beta}\eta_{\beta}\left\{ \tilde{D},\bar{A}\right\}
-{\sqrt{2}i\over2}C^{\alpha\beta}(\sigma^{\mu\nu}\eta)_\beta
\left\{ F_{\mu\nu}, \bar{A}\right\}
-{\sqrt{2}g\over2} C^{\alpha\beta}\eta_{\beta}
\left\{ \bar{A}, [\bar{A},A]\right\}
\nonumber\\
&&+{\sqrt{2}\over4}
\det C
\left(\{\bar{\lambda}\bar{\lambda},\bar{A}\}
+2\bar{\lambda}_{\dot{\alpha}}\bar{A}\bar{\lambda}^{\dot{\alpha}}\right)
\eta^{\alpha},
\nonumber\\
\delta_\eta\bar{\lambda}&=&\sqrt{2}i\bar{\sigma}^{\mu}\eta D_{\mu}\bar{A},
\nonumber\\
\delta_\eta \tilde{D}&=&
-\eta \sigma^{\mu} D_{\mu} \bar{\psi}-\sqrt{2}g[\eta\lambda, \bar{A}]
-{\sqrt{2}\over2} i C^{\alpha\beta}\eta_\beta
D_{\mu}\left\{ \bar{A}, (\sigma^{\mu}\bar{\lambda})_\alpha\right\}
-ig C^{\alpha\beta}\eta_\beta
\left\{\bar{A}, [\bar{A}, \psi_\alpha]\right\} ,
\nonumber\\
\delta_\eta A&=& \sqrt{2}\eta\lambda
+ i  C^{\alpha\beta}\eta_\beta
\left\{\psi_\alpha, \bar{A}\right\} ,
\nonumber\\
\delta_\eta \psi^\alpha&=& i\eta^\alpha
\tilde{D}+ig\eta^\alpha[A,\bar{A}]
-\varepsilon^{\alpha\beta}(\sigma^{\mu\nu}\eta)_\beta F_{\mu\nu}
-i C^{\alpha\beta}\eta_{\beta}
\left\{(\bar{\lambda}\bar{\lambda})-\left\{\bar{A}, F\right\} \right\} ,
\nonumber\\
\delta_\eta F&=& i\sqrt{2}\eta\sigma^{\mu} D_{\mu}\bar{\lambda}
+2gi [\bar{A},\eta\psi] ,
\nonumber\\
\delta_\eta \bar{A}&=&0 ,
\nonumber\\
\delta_\eta\bar{\psi}_{\dot{\alpha}}&=& 
C^{\alpha\beta}\eta_\beta \sigma^{\mu}_{\alpha\dot{\alpha}}
\left\{\bar{A}, D_{\mu}\bar{A}\right\} ,
\nonumber\\
\delta_\eta \bar{F}&=& 
\sqrt{2}g
C^{\alpha\beta}\eta_{\beta} \left\{\bar{A}, 
[\bar{A},\lambda_\alpha]\right\} 
+
{\sqrt{2}i\over 4}\det C
\biggl[3\left\{\bar{A}, 
\left\{ \eta\sigma^{\mu}\bar{\lambda}, D_{\mu}\bar{A}\right\}\right\}
\nonumber\\
&&+
2 D_{\mu}\bar{A}\bar{A}\eta\sigma^{\mu}\bar{\lambda}
+
2 \eta\sigma^{\mu}\bar{\lambda}\bar{A}D_{\mu}\bar{A}
+
2\left\{\bar{A}, 
\left\{ \eta\sigma^{\mu}D_{\mu}\bar{\lambda}, \bar{A}\right\}\right\}
\biggr],
\label{eq:dsusyeta1}
\end{eqnarray}
where we have written down only the part of ${\cal N}=(1,0)$ supersymmetry.

We now compute the Noether currents associated with deformed
${\cal N}=(1,0)$ supersymmetry transformations 
$\delta_{\xi}$ and $\delta_{\eta}$. 
Let $X^{\mu}_{\xi}$ be the total derivative term obtained from 
 the variation of the
Lagrangian associated with the transformation $\delta_{\xi}$:
$$
\delta_{\xi}{\cal L}=\partial_{\mu} X^{\mu}_{\xi}.
$$
Then the supercurrent $N^{\mu}_{1\alpha}$ is defined by 
\begin{equation}
 \xi^{\alpha} N^{\mu}_{1\alpha}
={\partial {\cal L}\over \partial (\partial_{\mu} \varphi_{A})}\delta_{\xi} \varphi_{A}
-X_{\xi}^{\mu}
\end{equation}
where $\varphi_{A}$ are component fields in the WZ gauge.
The other supercurrent $N^{\mu}_{2\alpha}$ associated with the transformation
 $\delta_\eta$ is defined in a similar way.
{}From the Lagrangian (\ref{eq:lag2a}) and the transformations
(\ref{eq:tra1}), we get
\begin{eqnarray}
 \xi N^{\mu}_1&=&{1\over k}{\rm tr}
\Bigl\{
-i(F^{\mu\nu}+\tilde{F}^{\mu\nu})\xi\sigma_{\nu}\bar{\lambda}
+\sqrt{2}D_{\nu}\bar{A}\xi\sigma^{\nu}\bar{\sigma}^{\mu}\psi
+g \xi \sigma^{\mu}\bar{\lambda}[A,\bar{A}]
\nonumber\\
&&
+(\xi\sigma_{\nu} \bar{\lambda}) C^{\mu\nu}\bar{\lambda}\bar{\lambda}
-(\xi\sigma_{\nu} \bar{\lambda}) C^{\mu\nu}\left\{\bar{A},F\right\}
\Bigr\}.
\label{eq:curr1}
\end{eqnarray}
The supercurrent $N^{\mu}_{2}$  is given by
\begin{eqnarray}
\eta N^{\mu}_2&=& {1\over k}{\rm tr}\Bigl\{
i(F^{\mu\nu}+\tilde{F}^{\mu\nu})\eta\sigma_{\nu}\bar{\psi}
+\sqrt{2}D_{\nu}\bar{A}\eta\sigma^{\nu}\bar{\sigma}^{\mu}\lambda
-g \eta\sigma^{\mu}\bar{\psi}[A,\bar{A}]
\nonumber\\
&&
-{\sqrt{2}\over2} C^{\alpha\beta}
\left\{F^{\mu\nu}+\tilde{F}^{\mu\nu}, \bar{A}\right\}
\eta_\alpha (\sigma_{\nu}\bar{\lambda})_\beta
-C^{\mu\nu}\eta\sigma_{\nu}\bar{\lambda}
\left(\bar{\lambda}\bar{\lambda}-\left\{\bar{A},F\right\}\right)
\nonumber\\
&&
+iC^{\alpha\beta} \left\{\bar{A}, D_{\nu}\bar{A}\right\}
\eta_\alpha (\sigma^{\nu} \bar{\sigma}^{\mu}\psi)_\beta
+ig {\sqrt{2}\over2}C^{\mu\nu}\eta\sigma_{\nu}\bar{\lambda}
\left\{ \bar{A},[\bar{A},A]\right\}
\nonumber\\
&&
-i{\sqrt{2}\over2}
\det C\eta\sigma^{\mu}\bar{\lambda}
\left(
\left\{\bar{A},\bar{\lambda}\bar{\lambda}\right\}
-\left\{\bar{A},\left\{\bar{A},F\right\}\right\}
\right)
\Bigr\},
\label{eq:curr2}
\end{eqnarray}
which contains $O(C^2)$ corrections.
For $C=0$, we recover the undeformed supercurrents \cite{WiOl,AlHa}.
The supercharge $Q_{i\alpha}$ is defined by
$$
Q_{i\alpha}=\int d^3x N_{i\alpha}(x).
$$
We now examine the anticommutation relations for supercharges $Q_{i\alpha}$.
We will use the equal-time anticommutation relations for fermions
\begin{equation}
 \left\{\psi_{\alpha}(x),\bar{\psi}_{\dot{\alpha}}(y)\right\}
=\delta_{\alpha\dot{\alpha}}\delta^3(x-y),\quad
 \left\{\lambda_{\alpha}(x),\bar{\lambda}_{\dot{\alpha}}(y)\right\}
=\delta_{\alpha\dot{\alpha}}\delta^3(x-y).
\label{eq:etac1}
\end{equation}
{}From (\ref{eq:curr1}), (\ref{eq:curr2}) and (\ref{eq:etac1}), we 
find that $\left\{Q_{1\alpha}, Q_{1\beta}\right\}$ and 
$\left\{Q_{1\alpha}, Q_{2\beta}\right\}$ are undeformed:
\begin{eqnarray}
\left\{ Q_{1\alpha}, Q_{1\beta}\right\}
&=&0,
\label{eq:ac0}
\\
\left\{ Q_{1\alpha}, Q_{2\beta}\right\}
&=&
2\sqrt{2}i\varepsilon_{\alpha\beta}
\int d^3 x {1\over k}{\rm tr}
\Bigl[
(F_{0\ell}+\tilde{F}_{0\ell})D^{\ell}\bar{A}
\Bigr].
\label{eq:ac1}
\end{eqnarray}
The r.h.s. of (\ref{eq:ac1})
comes from the 1st and 2nd terms in (\ref{eq:curr1}) and
(\ref{eq:curr2}) 
and we have eliminated auxiliary fields by using the equations of motion.
Eq. (\ref{eq:ac1}) is nothing but the central charge obtained by Witten and
Olive \cite{WiOl}.

The $C$-deformation arises in the anticommutation relation
$\left\{Q_{2\alpha}, Q_{2\beta}\right\}$, which 
 is given by
\begin{equation}
 \left\{Q_{2\alpha}, Q_{2\beta}\right\}
=4 C_{\alpha\beta} 
\int d^3 x {1\over k}{\rm tr}
\Bigl[
(F_{0\ell}+\tilde{F}_{0\ell})D^{\ell}\bar{A}^2
\Bigr].
\label{eq:ac2}
\end{equation}
The r.h.s. of (\ref{eq:ac2}) is obtained  from the anticommutation relation
among the 1st, 2nd, 
4th and 7th terms in the current (\ref{eq:curr2}). 
Eq. (\ref{eq:ac2}) gives still the topological charge but its dependence on
the vacuum
expectation value of the Higgs fields is different from the undeformed
topological charge (\ref{eq:ac1}).

In this paper, we will further study the deformed ${\cal N}=(1,0)$ 
supersymmetry algebra for generic deformation case.
In order to study deformed theories in extended superspace, 
it is convenient to introduce non(anti)commutative $\mathcal{N}=2$ 
harmonic superspace \cite{GaIvOgSo} with supercoordinates
$(x^{\mu},\theta_{i}^{\alpha},\bar{\theta}^{\dot{\alpha} i},u^{\pm i})$.
Here $i=1,2$ are $SU(2)_{R}$ $R$-symmetry group 
indices. $u^{\pm i}$ are the harmonic variables.
The non(anti)commutativity in $\mathcal{N}=2$ harmonic superspace 
\cite{IvLeZu} is introduced by 
\begin{eqnarray}
&&
[x_{L}^{\mu},x_{L}^{\nu}]_{\ast}=[x_{L}^{\mu},\theta_{i}^{\alpha}]_{\ast}=
[x_{L}^{\mu},\bar{\theta}^{\dot{\alpha} i}]_{\ast}=0, 
\nonumber\\
&&
\{\bar{\theta}^{\dot{\alpha} i},\bar{\theta}^{\dot{\beta} j}\}_{\ast}=
\{\bar{\theta}^{\dot{\alpha} i},\theta^{\alpha}_{j}\}_{\ast}=0,
\quad 
\{\theta_{i}^{\alpha},\theta_{j}^{\beta}\}_{\ast}=
C^{\alpha\beta}_{ij}.
\label{nac2}
\end{eqnarray}
where $C^{\alpha\beta}_{ij}$ is the deformation parameter. 
$x_{L}^{\mu}\equiv x^{\mu}+i\theta_{i}\sigma^{\mu}\bar{\theta}^{i}$ 
is the $\mathcal{N}=2$ chiral coordinates. 
The non(anti)commutativity (\ref{nac2}) is realized by using 
the $\ast$-product:
\begin{equation}
f \ast g(\theta)=f(\theta)
\exp\left(-\frac{1}{2}\overleftarrow{Q^{i}_{\alpha}}
C^{\alpha\beta}_{ij}
\overrightarrow{Q^{j}_{\beta}}\right)
g(\theta).
\label{star}
\end{equation}
where $Q^i_\alpha$ are supersymmetry generators 
which act on the $\mathcal{N}=2$ superspace.
The deformation parameter $C^{\alpha\beta}_{ij}$ has a symmetric property 
$C^{\alpha\beta}_{ij}=C^{\beta\alpha}_{ji}$ and can be decomposed as 
\begin{equation}
C^{\alpha\beta}_{ij}
=C^{\alpha\beta}_{(ij)}+\frac{1}{4}\epsilon_{ij}\varepsilon^{\alpha\beta}C_{s}.
\end{equation}
Here $C_{s}$ corresponds to the singlet deformation and
$A_{(ij)}$ denotes the symmetrized sum of $A_{ij}$ over indices $i$ and $j$.

At present the full component action for ${\cal N}=2$ $U(N)$ gauge theory
is not yet constructed for generic deformation parameters except for
the singlet deformation case\cite{FeIvLeSoZu}.
Recently, the exact form of the bosonic action of $U(1)$ theory is computed 
in \cite{DeIvLeQu} for the particular type of 
the non-singlet deformation
$C^{\alpha\beta}_{(ij)}=c^{\alpha\beta}b_{(ij)}$.
In the present work we will construct the supercharges
for the $O(C)$ action of ${\cal N}=2$ $U(1)$ gauge theory \cite{ArItOh2}.
Although the supersymmetry algebra does not have central extension 
due to the absence of scalar potential, 
we would expect that the similar algebraic structure also appear in the 
$U(N)$ case.

The $O(C)$ Wess-Zumino gauge 
Lagrangian of the $U(1)$ theory \cite{ArItOh2} in the
deformed ${\cal N}=2$ harmonic superspace is given by  
\begin{eqnarray}
\mathcal{L}
&=&
-\frac{1}{4}(1+\sqrt{2}C_{s}\bar{\phi})
F_{\mu\nu}(F^{\mu\nu}+\tilde{F}^{\mu\nu})
\nonumber\\
&&\ \ 
-i(1-\frac{1}{\sqrt{2}}C_{s}\bar{\phi})
\psi^{i}\sigma^{\mu}\partial_{\mu}\bar{\psi}_{i}
+\frac{i}{\sqrt{2}}C_{s}\partial_{\mu}\bar{\phi}
(\psi^{i}\sigma^{\mu}\bar{\psi}_{i})
\nonumber\\
&&\ \ 
+\phi\partial^{2}\bar{\phi}
+\frac{1}{4}(1-\sqrt{2}C_{s}\bar{\phi})D^{ij}D_{ij}
\nonumber\\
&&\ \ 
-2\sqrt{2}iC^{\alpha\beta}_{(ij)}\psi^{i}_{\alpha}
(\sigma^{\mu}\bar{\psi}^{j})_{\beta}\partial_{\mu}\bar{\phi}
-\frac{2\sqrt{2}}{3}iC^{\alpha\beta}_{(ij)}\psi^{i}_{\alpha}
(\sigma^{\mu}\partial_{\mu}\bar{\psi}^{j})_{\beta}\bar{\phi}
\nonumber\\
&&\ \ 
+\frac{i}{2}C_{s}\bar{\psi}^{i}\bar{\psi}^{j}D_{ij}
-iC^{\mu\nu}_{(ij)}\bar{\psi}^{i}\bar{\psi}^{j}F_{\mu\nu}
+\frac{1}{\sqrt{2}}C^{\mu\nu}_{(ij)}D^{ij}F_{\mu\nu}\bar{\phi}+O(C^{2}).
\label{lag}
\end{eqnarray}
We can compute the $O(C)$ contributions of 
$C^{\alpha\beta}_{(ij)}$ and $C_{s}$ to the deformed supersymmetry 
separately. 
We firstly consider the case of non-singlet deformation $C_{s}=0$. 
Setting $C_{s}=0$, the Lagrangian (\ref{lag}) becomes
\begin{eqnarray}
\mathcal{L}^{non-singlet}
&=&
-\frac{1}{4}F_{\mu\nu}(F^{\mu\nu}+\tilde{F}^{\mu\nu})
+\phi\partial^{2}\bar{\phi}-i\psi^{i}\sigma^{\mu}\partial_{\mu}\bar{\psi}_{i}
+\frac{1}{4}D^{ij}D_{ij}
\nonumber\\
&&\ \ 
-2\sqrt{2}iC^{\alpha\beta}_{(ij)}\psi^{i}_{\alpha}
(\sigma^{\mu}\bar{\psi}^{j})_{\beta}\partial_{\mu}\bar{\phi}
-\frac{2\sqrt{2}}{3}iC^{\alpha\beta}_{(ij)}\psi^{i}_{\alpha}
(\sigma^{\mu}\partial_{\mu}\bar{\psi}^{j})_{\beta}\bar{\phi}
\nonumber\\
&&\ \ 
-iC^{\mu\nu}_{(ij)}\bar{\psi}^{i}\bar{\psi}^{j}F_{\mu\nu}
+\frac{1}{\sqrt{2}}C^{\mu\nu}_{(ij)}D^{ij}F_{\mu\nu}\bar{\phi}+O(C^{2}).
\label{nslag}
\end{eqnarray}
The $\mathcal{N}=(1,0)$ supersymmetry transformation law in the WZ gauge is 
given in \cite{ArItOh4} by
\begin{eqnarray}
\tilde{\delta}_{\xi}\phi
&=&
-\sqrt{2}\xi^{i}\psi_{i}-\frac{8}{3}i(\xi^{i}\varepsilon 
C_{(ij)}\psi^{j})\bar{\phi}+O(C^{2}),
\nonumber\\
\tilde{\delta}_{\xi}\bar{\phi}
&=&
0,
\nonumber\\
\tilde{\delta}_{\xi}A_{\mu}
&=&
i\xi^{i}\sigma_{\mu}\bar{\psi}_{i}+2\sqrt{2}i
(\xi^{i}\varepsilon C_{(ij)}\sigma_{\mu}\psi^{j})\bar{\phi}+O(C^{2}),
\nonumber\\
\tilde{\delta}_{\xi}\psi^{\alpha i}
&=&
-(\xi^{i}\sigma^{\mu\nu})^{\alpha}F_{\mu\nu}
-D^{ij}\xi_{j}^{\alpha}-i(\xi^{i}\sigma_{\mu\nu})^{\alpha}
C^{\mu\nu}_{(jk)}\bar{\psi}^{j}\bar{\psi}^{k}
+2\sqrt{2}D^{(ij}(\xi^{k)}\varepsilon C_{(jk)})^{\alpha}\bar{\phi}
\nonumber\\
&&
-\bigl\{2\sqrt{2}(\xi^{j}\varepsilon C_{(jk)}\sigma^{\mu\nu})^{\alpha}
+\frac{2\sqrt{2}}{3}(\xi^{j}\sigma^{\mu\nu}\varepsilon C_{(jk)})^{\alpha}
+\sqrt{2}C^{\mu\nu}_{(jk)}\xi^{\alpha j}\bigr\}\epsilon^{ki}\bar{\phi}
F_{\mu\nu}+O(C^{2}),
\nonumber\\
\tilde{\delta}_{\xi}\bar{\psi}^{i}_{\dot{\alpha}}
&=&
\sqrt{2}(\xi^{i}\sigma^{\mu})_
{\dot{\alpha}}\partial_{\mu}\bar{\phi}+2(\xi^{j}\varepsilon C_{(jk)}
\sigma^{\mu})_{\dot{\alpha}}\partial_{\mu}\bar{\phi}^{2}\epsilon^{ki}+O(C^{2}),
\nonumber\\
\tilde{\delta}_{\xi}D^{ij}
&=&
-2i\xi^{(i}\sigma^{\mu}\partial_{\mu}\bar{\psi}^{j)}
-6\sqrt{2}i\epsilon^{k(l}\partial_{\mu}\{(\xi^{i}\varepsilon C_{(kl)}
\sigma^{\mu}\bar{\psi}^{j)})\bar{\phi}\}
\nonumber\\
&&
+2\sqrt{2}i\epsilon^{il}\epsilon^{jm}(\xi^{k}\varepsilon C_{(lm)}
\sigma^{\mu}\bar{\psi}_{k})\partial_{\mu}\bar{\phi}+O(C^{2}).
\label{nstransf}
\end{eqnarray}
Under the supersymmetric transformation (\ref{nstransf}), 
the Lagrangian (\ref{nslag}) is invariant up to 
$O(C)$. Then the supercurrent is 
\begin{eqnarray}
\xi^{i}\tilde{N}_{i}^{\mu}
&=&
-i(F^{\mu\nu}+\tilde{F}^{\mu\nu})\xi^{i}\sigma_{\nu}\bar{\psi}_{i}
-\sqrt{2}i(\psi'_{i}\sigma^{\mu}\bar{\sigma}^{\nu}\xi^{i})\partial_{\nu}
\bar{\phi}
\nonumber\\
&&
-(2C^{\mu\nu}_{(ij)}\bar{\psi}^{i}\bar{\psi}^{j}+\sqrt{2}iC^{\mu\nu}_{(ij)}
D^{ij}\bar{\phi})\xi^{k}\sigma_{\nu}\bar{\psi}_{k}
\nonumber\\
&&
-2\sqrt{2}i(F^{\mu\nu}+\tilde{F}^{\mu\nu})\xi^{i}\varepsilon C_{(ij)}
\sigma_{\nu}\bar{\psi}^{j}\bar{\phi}
-2i\xi^{i}\varepsilon C_{(ij)}\sigma^{\nu}\bar{\sigma}^{\mu}\psi'^{j}
\partial_{\nu}\bar{\phi}^{2}+O(C^{2}),
\label{nscurr}
\end{eqnarray}
where $\psi'$ is defined by 
$
\psi^{\prime\alpha}_{i}=\psi^{\alpha}_{i}-\frac{2\sqrt{2}}{3}
C^{\alpha\beta}_{(ij)}\bar{\phi}\psi_{\beta}^{j}.
$
Using the anticommutation relation 
$\{\psi^{\prime i}_{\alpha}(x),\bar{\psi}_{j\dot{\beta}}(y)\}=
\delta^{i}_{j}\delta_{\alpha\dot{\beta}}\delta^{3}(x-y)$, 
we find the deformation of the central charge as 
\begin{eqnarray}
\{Q_{i\alpha},Q_{j\beta}\}
&=&
2\sqrt{2}\epsilon_{ij}\varepsilon_{\alpha\beta}
\!\int\! d^{3}x\, (F^{0\ell}+\tilde{F}^{0\ell})\partial_{\ell}\bar{\phi}
\nonumber\\
&&\ \ \ \ 
+8C_{(ij),\alpha\beta}\int\! d^{3}x\,(F^{0\ell}+\tilde{F}^{0\ell})
\partial_{\ell}\bar{\phi}^{2}+O(C^{2}).
\label{nsalg}
\end{eqnarray}
The algebra (\ref{nsalg}) coincides with the $U(1)$ case of 
(\ref{eq:ac0})--(\ref{eq:ac2}) 
under the reduction of the deformation parameter 
$C^{\alpha\beta}_{(ij)}=C^{\alpha\beta}_{11}\delta^{1}_{i}\delta^{1}_{j}$. 

We note that 
in the case of singlet deformation $C^{\alpha\beta}_{(ij)}=0$, $C_{s}\neq 0$, 
the exact form of the Lagrangian is obtained in \cite{FeIvLeSoZu,ArIt3}
such as 
\begin{equation}
\mathcal{L}^{singlet}=\biggl(1+\frac{1}{\sqrt{2}}C_{s}\bar{\phi}\biggr)^{2}
\biggl[-\frac{1}{4}F_{\mu\nu}(F^{\mu\nu}+\tilde{F}^{\mu\nu})
+\phi\partial^{2}\bar{\phi}-i\psi^{i}\sigma^{\mu}\partial_{\mu}\bar{\psi}_{i}
+\frac{1}{4}D^{ij}D_{ij}\biggr].
\label{slag}
\end{equation}
Here the component fields in (\ref{slag}) except for $A_{\mu}$ and 
$\bar{\phi}$ are redefined from those in (\ref{lag}) 
so that the $\mathcal{N}=(1,0)$ supersymmetry transformation law 
in the WZ gauge is the same as the undeformed one\cite{FeIvLeSoZu,ArIt3}: 
\begin{eqnarray}
&&
\hat{\delta}_{\xi}A_{\mu}=i\xi^{i}\sigma_{\mu}\bar{\psi}_{i}, \quad 
\hat{\delta}_{\xi}\phi=-\sqrt{2}\xi^{i}\psi_{i}, \quad 
\hat{\delta}_{\xi}\bar{\phi}=0,
\nonumber\\
&&
\hat{\delta}_{\xi}\psi^{\alpha i}=-(\xi^{i}\sigma^{\mu\nu})^{\alpha}F_{\mu\nu}
-D^{ij}\xi_{j}^{\alpha}, \quad 
\hat{\delta}_{\xi}\bar{\psi}^{i}_{\dot{\alpha}}=\sqrt{2}(\xi^{i}\sigma^{\mu})_
{\dot{\alpha}}\partial_{\mu}\bar{\phi},\nonumber\\
&&
\hat{\delta}_{\xi}D^{ij}=-i(\xi^{i}\sigma^{\mu}\partial_{\mu}\bar{\psi}^{j}
+\xi^{j}\sigma^{\mu}\partial_{\mu}\bar{\psi}^{i}).
\label{stransf}
\end{eqnarray}
The supercurrent generating the transformation (\ref{stransf}) is given by
\begin{equation}
\xi^{i}\hat{N}_{i}^{\mu}=
\biggl(1+\frac{1}{\sqrt{2}}C_{s}\bar{\phi}\biggr)^{2}
\Bigl[-i(F^{\mu\nu}+\tilde{F}^{\mu\nu})\xi^{i}\sigma_{\nu}\bar{\psi}_{i}
-\sqrt{2}i(\psi_{i}\sigma^{\mu}\bar{\sigma}^{\nu}\xi^{i})\partial_{\nu}
\bar{\phi}\Bigr].
\label{scurr}
\end{equation}
{}From (\ref{scurr}), we get the supersymmetry algebra as
\begin{equation}
\{Q_{i\alpha},Q_{j\beta}\}=
2\sqrt{2}\epsilon_{ij}\varepsilon_{\alpha\beta}
\!\int\! d^{3}x\, \biggl(1+\frac{1}{\sqrt{2}}C_{s}\bar{\phi}\biggr)^{2}
(F^{0\ell}+\tilde{F}^{0\ell})\partial_{\ell}\bar{\phi}.
\label{salg}
\end{equation}
Since $A_{\mu}$ and $\bar{\phi}$ are not redefined, 
we can directly combine (\ref{nsalg}) and (\ref{salg}) 
into the central charge formula. 
Then we finally obtain 
the supersymmetry algebra in generic deformation case:
\begin{eqnarray}
\{Q_{i\alpha},Q_{j\beta}\}
&=&
2\sqrt{2}\epsilon_{ij}\varepsilon_{\alpha\beta}
\!\int\! d^{3}x\, (F^{0\ell}+\tilde{F}^{0\ell})\partial_{\ell}\bar{\phi}
\nonumber\\
&&\ \ 
+8C_{ij,\alpha\beta}\int\! d^{3}x\,(F^{0\ell}+\tilde{F}^{0\ell})\partial_{\ell}
\bar{\phi}^{2}+O(C^{2}).
\label{alg}
\end{eqnarray}
The $O(C^{2})$ term of r.h.s. in (\ref{alg}) is generally nonzero 
as we have seen in the case of the singlet deformation (\ref{salg}).

We also examine the the supersymmetry algebra 
in $\mathcal{N}=2$ supersymmetric $U(N)$ gauge theory 
with singlet deformation. 
The Lagrangian $\mathcal{L}^{singlet}_{U(N)}$ obtained in \cite{FeIvLeSoZu} 
is of the form 
\begin{eqnarray}
\mathcal{L}^{singlet}_{U(N)}
&=&
\frac{1}{k}{\rm tr}\Biggl[-\frac{1}{4}(L^{2}F_{\mu\nu})F^{\mu\nu}
-\frac{1}{4}(L^{2}F_{\mu\nu})\tilde{F}^{\mu\nu}
-(L^{2}\psi^{i})\sigma^{\mu}D_{\mu}\bar{\psi}_{i}
+(L^{2}\phi)D^{2}\bar{\phi}
\nonumber\\
&&\ \ 
+\frac{1}{4}(L^{2}D_{ij})D^{ij}
-\frac{g}{\sqrt{2}}(L^{2}\psi^{i})[\bar{\phi},\psi_{i}]
+\frac{g}{\sqrt{2}}\bar{\psi}^{i}[L^{2}\phi,\bar{\psi}_{i}]
-\frac{g^{2}}{2}(L[\phi,\bar{\phi}])^{2}\Biggr]
\nonumber\\
&&\ \ 
+(\mbox{higher-derivative terms}),
\label{nasing}
\end{eqnarray}
where the operator $L$ is defined by 
\begin{equation}
L=1+\frac{C_{s}}{2\sqrt{2}}\{\bar{\phi},\,\cdot\,\}.
\end{equation}
The higher-derivative terms in the Lagrangian (\ref{nasing}) can be absorbed 
by suitable field redefinitions\cite{FeIvLeSoZu}. 
After these field redefinitions, as in the $U(1)$ case, 
the Lagrangian (\ref{nasing}) is invariant up to the total derivative
under the undeformed $\mathcal{N}=(1,0)$ supersymmetry transformations: 
\begin{eqnarray}
\delta_{\xi}A_{\mu}
&=&
i\xi^{i}\sigma_{\mu}\bar{\psi}_{i},\quad 
\delta_{\xi}\phi=-\sqrt{2}\xi^{i}\psi_{i}, \quad 
\delta_{\xi}\bar{\phi}=0,
\nonumber\\
\delta_{\xi}\psi_{i}
&=&
\sigma^{\mu\nu}\xi_{i}F_{\mu\nu}+D_{ij}\xi^{j}
-ig\xi_{i}[\phi,\bar{\phi}],\quad 
\delta_{\xi}\bar{\psi}_{i}=-\sqrt{2}\bar{\sigma}^{\mu}\xi_{i}D_{\mu}\bar{\phi},
\nonumber\\
\delta_{\xi}D_{ij}
&=&
-2i\{\xi_{(i}\sigma^{\mu}D_{\mu}\bar{\psi}_{j)}
+\sqrt{2}g[\xi_{(i}\psi_{j)},\bar{\phi}]\}.
\label{nastransf}
\end{eqnarray}
The variation of the Lagrangian becomes
\begin{equation}
\delta_{\xi}\mathcal{L}^{singlet}_{U(N)}=\partial_{\mu}X^{\mu}, \quad 
X^{\mu}=-\frac{1}{k}{\rm tr}\Bigl(g\xi^{i}\sigma^{\mu}\bar{\psi}_{i}
[L^{2}\phi,\bar{\phi}]\Bigr).
\end{equation}
Here we used the formula $L[\bar{\phi},U]=[\bar{\phi},LU]$ for an arbitrary 
field $U$. The supercurrent is given by
\begin{eqnarray}
\xi^{i}N^{\mu}_{i}
&=&
\frac{1}{k}{\rm tr}\Bigl[-i\{L^{2}(F^{\mu\nu}+\tilde{F}^{\mu\nu})\}
\xi^{i}\sigma_{\nu}\bar{\psi}_{i}
\nonumber\\
&&\ \ \ 
-\sqrt{2}i\{(L^{2}\psi_{i})\sigma^{\mu}\bar{\sigma}^{\nu}\xi^{i}\}
D_{\nu}\bar{\phi}
+g\xi^{i}\sigma^{\mu}\bar{\psi}_{i}
[L^{2}\phi,\bar{\phi}]\Bigr].
\label{nascurr}
\end{eqnarray}
We then obtain the deformation of the central charge as 
\begin{equation}
\{Q_{i\alpha},Q_{j\beta}\}=
2\sqrt{2}\epsilon_{ij}\varepsilon_{\alpha\beta}
\!\int\! d^{3}x\, \frac{1}{k}{\rm tr}
\Bigl[\{L^{2}(F^{0\ell}+\tilde{F}^{0\ell})\}
D_{\ell}\bar{\phi}\Bigr].
\label{nasalg}
\end{equation}
The algebra (\ref{nasalg}) is reduced to (\ref{salg}) 
in the case of $U(1)$ gauge group.

In this paper we have studied the central extension of the deformed
${\cal N}=(1,0)$ supersymmetry algebra.
For $U(N)$ gauge group, we have obtained the $C$-deformed 
central charge in the cases of the  deformed ${\cal N}=1$ superspace and
${\cal N}=2$ harmonic superspace with the singlet deformation.
In generic deformation case, we have computed the $O(C)$-correction to
the central charge for the $U(1)$ gauge group. 
It is important to study the full action of the deformed ${\cal N}=2$ 
theory in order to discuss the complete $C$-deformed central charges.
It is an interesting problem to find monopole and
dyon solutions and study how the BPS structure is modified by the
non(anti)commutativity. 
It is also interesting to study physical effects of this
non(anti)commutativity in the strong coupling region of the 
theory.

\subsection*{Acknowledgements}
H. N. is supported by Grant-in-Aid for Scientific Research from 
the Japanese Ministry of Education and Science, 
No. 16028203 for the priority area ``origin of mass".

\end{document}